\documentclass[a4paper,12pt,leqno]{article}

\usepackage{graphicx}
\usepackage{subeqn}
\usepackage{amssymb}
\usepackage{lineno}
\usepackage{bm}
\usepackage{array}
\usepackage{color}
\usepackage{subfigure}

\date{ }

\begin{document}

\title{Fractional Spectral Moments for Digital Simulation of Multivariate Wind Velocity Fields
\footnote{Publication info: Cottone G., Di Paola M., Fractional spectral moments for digital simulation of multivariate wind velocity fields, Journal of Wind Engineering and Industrial Aerodynamics, Volume 99, Issues 6–7, June–July 2011, Pages 741-747, ISSN 0167-6105, 10.1016/j.jweia.2011.03.006.} }

\author{Giulio Cottone $^{1,2}$ \footnote{E-mail: giulio.cottone@tum.de; giuliocottone@yahoo.it} and Mario Di Paola $^{1}$ \footnote{E-mail: mario.dipaola@unipa.it}\\
\small $^1$ Dipartimento di Ingegneria Civile, Aerospaziale ed Ambientale, \\ 
\small Universit\'{a} degli Studi di Palermo, Viale delle Scienze, 90128 Palermo, Italy\\
\small $^2$ Engineering Risk Analysis Group, Technische Universit\"at M\"unchen,\\
\small Theresienstr.90, building N6, 80290, Germany
}


\maketitle


\begin{abstract}
In this paper, a method for the digital simulation of wind velocity fields by Fractional Spectral Moment function is proposed. It is shown that by constructing a digital filter whose coefficients are the fractional spectral moments, it is possible to simulate samples of the target process as superposition of Riesz fractional derivatives of a Gaussian white noise processes. The key of this simulation technique is the generalized Taylor expansion proposed by the authors. The method is extended to multivariate processes and practical issues on the implementation of the method are reported.
\end{abstract}

\section{Introduction} 
Digital simulation of wind velocity field is needed in the design of wind exposed structures. The literature on this topic is vast and we refer to the review of Kareem, 2006 which provides a synthetic overview of different possible approaches. In this paper we confine our attention on the simulation of wind velocity fields which are both Gaussian and stationary. Under these assumptions, two strategies for the digital generation of wind velocity samples  became standard: i) by superposition of trigonometric functions (Borgman, 1969; Shinozuka, 1971); ii) by digital filtering technique, which consists in calibrating the output of linear differential equation, called filter, excited by a white noise process. The latter are commonly referred to as auto-regressive (AR) algorithms, moving average (MA) and auto-regressive moving-average (ARMA) algorithms (see Deodatis and Shinozuka, 1988; Deodatis, 1995; Kozin, 1988; Naganuma et al, 1987; Saramas et al, 1985; Spanos and Mignolet 1986).  

This paper show a novel method for the representation of wind velocity fields by digital filtering. Yet, in contrast to the classical use of filters, that are linear differential equations, we propose to use fractional differential equation, in which fractional derivatives appears. This is an absolute novelty in wind engineering and it will be shown that this method is easily applicable to multivariate fields. Our results rely on fractional calculus that is slowly becoming of interest in many engineering fields. For brevity's sake many concept on fractional operators will be given without further theory and readers are referred to the textbooks of Podlubny, 1998 and Samko et al. 1993 for introductive and authoritative treatments of such a topic.

The method here proposed is based on the recently published papers of Cottone and Di Paola, 2010a and Cottone et al, 2010b, which posed the theoretical basis for a novel representation of stationary Gaussian processes as output of a fractional stochastic differential filter.

In this paper, we extend our previous results both investigating on the digital simulation of samples with assigned statistics, useful in wind engineering, and applying the method to multivariate wind velocity field ${\bf{V}}\left(t\right)$ with assigned Power Spectral Density (PSD) matrix ${\bf{S}}_{\bf{V}} \left(\omega \right)$. 
The paper is organized as follow. In section 2, fundamentals on the representation of multivariate wind velocity fields is given. Due to the Gaussian assumption, only second order statistics are needed to characterize the wind velocity field. This will be done by specification of the PSD matrix ${\bf{S}}_{\bf{V}} \left(\omega \right)$, following the results given in  Solari and Piccardo, 2001.
Then, in sections 3-5, the representation method based on fractional spectral moments both for mono-variate and multivariate processes is given along with applications and implementation remarks, which aim to show the straightforward application of the theoretical concepts. 

\section{Probabilistic Description of Multivariate Wind Velocity Fields}
In this section the wind velocity field characterization is recalled. Readers are referred to the paper of Solari and Piccardo, 2001 and references therein reported, for a comprehensive treatment of the topic.
With the sake of introducing the notation, we consider that in the coordinate reference system $x, y, z$, the wind velocity process in a point $P\left(x,y,z\right)$ can be expressed as
\begin{equation} \label{1)} 
V_{P} \left(x,y,z;t\right)=\bar{V}\left(z\right)+V\left(x,y,z;t\right) 
\end{equation} 
in which the mean value $\bar{V}\left(z\right)$, thought as function of the sole elevation, is added to the fluctuation around the mean component $V\left(x,y,z;t\right)$ that is assumed to be a Gaussian stationary process with zero mean. The mean value follows the logarithmic profile $\bar{V}\left(z\right)=k^{-1} u_{*} \ln \left(z/z_{0} \right)$, where $k=0.4$ is the Karman's constant, $u_{*} $ is the shear velocity and $z_{0} $ is the roughness length. 

Given N points located in the space $P_{j} \left(x_{j} ,y_{j} ,z_{j} \right)$ with $j=1,2,...,N$ the wind velocity field can be represented as N-variate process by collecting the processes $V_{j} \left(x_{j} ,y_{j} ,z_{j} ;t\right)$ in the vector $
{\bf{V}}\left( t \right)=\left[ V_{1} V_{2} ...V_{N} \right]^{T} $. 
Due to the Gaussian assumption, ${\bf{V}}\left( t \right)$ is characterized by the power spectral density (PSD) matrix ${\bf{S}}_{\bf{V}} \left(\omega \right)$ that reads
\begin{equation} \label{2)} 
{\bf{S}}_{\bf{V}} \left(\omega \right)=\left[\begin{array}{cccc} {S_{V_{1} V_{1} } \left(\omega \right)} & {S_{V_{1} V_{2} } \left(\omega \right)} & {...} & {S_{V_{1} V_{N} } \left(\omega \right)} \\ {S_{V_{1} V_{2} } \left(\omega \right)} & {S_{V_{2} V_{2} } \left(\omega \right)} & {...} & {...} \\ {...} & {...} & {...} & {...} \\ {S_{V_{1} V_{N} } \left(\omega \right)} & {S_{V_{2} V_{N} } \left(\omega \right)} & {...} & {S_{V_{N} V_{N} } \left(\omega \right)} \end{array}\right] 
\end{equation} 
In this paper, we will assume that only the co-spectrum, i.e. the real part of ${\bf{S}}_{\bf{V}} \left(\omega \right)$, is needed to characterize the wind velocity field, neglecting the quad-spectrum, see Simiu \& Scalan, 1996. The terms of the PSD matrix are calculated following Solari \& Piccardo, 2001, in which the hypotheses of flat homogeneous terrain and near neutral atmospheric conditions are considered. Following the cited paper, explicit expression of the diagonal components of the PSD matrix reads
\begin{equation} \label{eq3} 
S_{V_{r} V_{r} } \left(\omega ,z\right)=\frac{d_{V} \, \sigma _{V}^{2} L_{V} \left(z\right)/\bar{V}\left(z\right)}{\left[1+1.5\, d_{V} \, \omega L_{V} \left(z\right)/\left(2\pi \bar{V}\left(z\right)\right)\right]^{5/3} }  
\end{equation} 
where $L_{V} $ is the integral length scale of turbulence that defines the position of the turbulence spectral content, $d_{V} =6.868$, $\sigma _{V}^{2} =\beta u_{*}^{2} $ is the variance of the velocity longitudinal component and $\beta $ is a non-dimensional coefficient defined as turbulence intensity factor. 

The cross-power spectral components of ${\bf{S}}_{\bf{V}}$ are expressed as 
\begin{equation} \label{4)} 
S_{V_{r} V_{s} } \left(\omega \right)=\sqrt{S_{V_{r} V_{r} } \left(\omega \right)S_{V_{s} V_{s} } \left(\omega \right)} \exp \left(-f_{rs} \left(\omega \right)\right) 
\end{equation} 
where
\begin{equation} \label{eq5} 
f_{rs} \left(\omega \right)=\frac{\left|\omega \right|\sqrt{C_{y}^{2} \left(y_{r} -y_{s} \right)^{2} +C_{z}^{2} \left(z_{r} -z_{s} \right)^{2} } }{2\pi \left(\bar{V}\left(z_{r} \right)+\bar{V}\left(z_{s} \right)\right)}  
\end{equation} 
being $C_{y} $ and $C_{z} $ two coefficient to be experimentally determined. Parameters can be estimated accordingly to Solari \& Piccardo, 2001. The method proposed in the next sections can also be applied to other form of the PSD matrix, but in this paper attention will be given to PSD constructed on eqs.(\ref{eq3})-(\ref{eq5}). 


As the spectral decomposition of the PSD matrix will be needed in the next section, we will briefly introduce some notation here for reference's sake and some concept on the use of such a decomposition for the representation of wind velocity fields. Let us indicate with ${\l} _{j} \left(\omega \right)$ the eigenvalue associated with the eigenvector ${\bf\psi }_{j} \left(\omega \right)$ of the matrix ${\bf{S}}_{\bf{V}} \left(\omega \right)$, ${\bf\Psi} \left(\omega \right)$ is the eigenmatrix, in which the jth column is the eigenvector ${\bf \psi }_{j} \left(\omega \right)$ and ${\bf L} \left(\omega \right)$ is the diagonal matrix whose jth element is $\l _{j} \left(\omega \right)$. Then, the relations
\begin{equation} \label{eq6} 
{\bf\Psi} ^{T} \left(\omega \right){\bf{S}}_{\bf{V}} \left(\omega \right){\bf{\Psi}} \left(\omega \right)={\bf{L}} \left(\omega \right) 
\end{equation} 
and 
\begin{equation} \label{7)} 
{\bf{\Psi}} ^{T} \left(\omega \right){\bf{\Psi}} \left(\omega \right)= {\bf{I}}
\end{equation} 
hold true, being ${\bf{I}}$ the identity matrix. The eigenmatrix decomposes the spectral density matrix such that
\begin{equation} \label{eq8} 
{\bf{S}}_{\bf{V}} \left(\omega \right) = {\bf{\Psi}}\left(\omega \right){\bf{\L}}\left(\omega \right){\bf{\Psi}} \left(\omega \right)^{T}  
\end{equation} 

The spectral decomposition has been used by different authors, see Di Paola, 1998 and Di Paola and Gullo, 2001, to express the wind field N-variate processes $V\left(t\right)$ in the form
\begin{equation} \label{eq9} 
{\bf{V}}\left(t\right)=\sum _{j=1}^{N}\int _{-\infty }^{\infty }{\it \psi }_{j} \left(\omega \right)\sqrt{\l _{j} \left(\omega \right)}  \, e^{i\omega t} dB_{j} \left(\omega \right)  
\end{equation} 
where $B_{j} \left(\omega \right)$ is a zero mean normal complex process having orthogonal increments, i.e. $E\left[dB_{j} \left(\omega \right)\right]=0$,  $dB_{j} \left(\omega \right)=d\bar{B}_{j} \left(\omega \right)$ and $E\left[dB_{j} \left(\omega _{r} \right)dB_{k}^{*} \left(\omega _{s} \right)\right]=\delta _{\omega _{r} \omega _{s} } \delta _{jk} d\omega _{r} $, having indicated by the over bar the complex conjugate and by $\delta _{jk} $ the Kronecker delta (i.e. $\delta _{jk} =0$ if  $j\ne k$ and $\delta _{jk} =1$ if  $j=k$). By eq.(\ref{eq9}) the vector process ${\bf{V}}\left(t\right)$ is decomposed into the sum of coherent and independent elementary vectors 
\begin{equation} \label{10)} 
Y_{j} \left(t\right)=\int _{-\infty }^{\infty }{\it \psi }_{j} \left(\omega \right)\sqrt{\l _{j} \left(\omega \right)}  \, e^{i\omega t} dB_{j} \left(\omega \right) 
\end{equation} 
such that 
\begin{equation} \label{11)} 
{\bf{V}}\left(t\right)=\sum _{j=1}^{N}Y_{j}(t)   
\end{equation} 
Moreover, retaining only the first $M\ll N$ most significant eigenvalues the latter is approximated in the form
\begin{equation} \label{eq12} 
{\bf{V}}\left(t\right)\cong \sum _{j=1}^{M}Y_{j}(t)  
\end{equation} 
As pointed out by Di Paola, 1998, although the spectral decomposition implies the calculation of the frequency dependent eigen-properties of the power spectral matrix ${\bf{S}}_{\bf{V}} \left(\omega \right)$, the computational pay off is that only $M\ll N$ independent component $Y_{j} $ can be considered. This concept can be combined also to the method here proposed as shown in the following. 

\section{Fractional Spectral Moments Decomposition}
In this section we start considering a Gaussian stationary mono-variate stochastic process with known statistics, while in the next section the proposed method is extended to Gaussian stationary multi-variate processes.
Let us assume we want to simulate a sample of the wind velocity in a point, with assigned PSD $S_V\left(\omega \right)$ whose form is given in the previous section. The method here proposed follows two steps. Firstly, we will express an assigned PSD function in terms of fractional spectral moments. Then, we will construct a digital filter whose output has the assigned PSD. It will be shown that fractional spectral moments of the linear system transfer function are the filter coefficients and that the filter has the form of a fractional differential equation. 
The hinge of the method is the generalized Taylor expansion introduced by the authors and applied in different contexts such as in probability, in Cottone and Di Paola, 2008, in stochastic dynamics by path integral solution in Cottone et al. 2009, in the solution of stochastic differential equations in Cottone et al. 2008. The application of the generalized Taylor expansion to representation of power spectral densities and correlation function has been introduced in Cottone and Di Paola, 2010, and Cottone et al. 2010 in the mono-variate case and readers are referred to those papers for more insight on the topic and relevant demonstrations. 

\subsection{Fractional moments for stochastic processes}
Let us consider a stationary Gaussian stochastic process $V(t)$ with target $S_V(\omega)$. Recall the definition of the Spectral Moments (SMs)  of the process, see Vanmarcke, 1972, defined as the integral
\begin{equation} \label{13)} 
\lambda _{V}^{j} =\int _{0}^{\infty }\omega ^{j}  G_{V} \left(\omega \right)d\omega \, \, \, \, \, \, \, \, \, \, \, \, \, \, \, \, \, \, \, \, \, \, \, \, \, \, \, \, \, \, \, \, j=0,1,2,... 
\end{equation} 
where $G_{V} \left(\omega \right)$ is the one-sided PSD calculated as $G_V(\omega) = 2 S_V(\omega)U(\omega)$, being $U(\omega)$ the unit step function. Physical meaning of SMs quantities has been given in Di Paola (1985). In Cottone and Di Paola, 2010a, the authors introduced a generalized class of SM, called the Fractional Spectral Moments (FSMs) of the one-sided PSD, that is 
\begin{equation} \label{eq14} 
\Lambda _{V} \left(\gamma \right)=\int _{0}^{\infty }\omega ^{\gamma }  G_{V} \left(\omega \right)d\omega \, \, \, \, \, \, \, \, \, \, \, \, \, \, \, \, \, \, \, \, \, \, \, \, \, \, \, \, \, \, \, \, \, \, \gamma \in {\mathbb C} 
\end{equation} 
Although the use of the adjective ``fractional'' to indicate moments that are of complex order $\gamma $ might be confusing, we keep it for similarity with the consolidated terminology of the Fractional Calculus which represent the calculus for derivation of real or complex order. 
The FSM function $\Lambda _{V} \left(\gamma \right)$ is a complex function that keeps all the information to restores both the power spectral density $S_{V} \left(\omega \right)$ and the correlation function $R_{V} \left(\tau \right)$. Skipping the details on the demonstration, see the paper above cited, we report here just the final formula needed in the following. To this aim, let us introduce the Riesz fractional integral and derivative as
\begin{subequations}\label{15)}
\begin{eqnarray} \label{15a)} 
{\left(I^{\gamma } f\right)\left(t\right)=\frac{1}{2\nu \left(\gamma \right)} \int _{-\infty }^{\infty }\frac{f\left(s \right)}{\left|t-s \right|^{1-\gamma } } {\rm d}s  \, \, \, \, \, \, \, \, \, \, \, \, \, \, \, \, \, \, \, \, \, \, \, \, \, \, \, \, Re\gamma >0,\, \gamma \ne 1,3,5,...}  
\end{eqnarray}
\begin{eqnarray} \label{15a)} 
{\left({\rm {\mathcal D}}^{\gamma } f\right)\left(t\right)=\frac{1}{2\nu \left(\gamma \right)} \int _{-\infty }^{\infty }\frac{f\left(t-s \right)-f\left(s \right)}{\left|t-s \right|^{1+\gamma } } {\rm d}s  \, \, \, \, \, \, \, \, \, Re\gamma >0,\, \gamma \ne 1,3,5,...} 
\end{eqnarray}
\end{subequations} 
where $\nu \left(\gamma \right)=\Gamma \left(\gamma \right)\cos \left(\pi \gamma /2\right)$, being $\Gamma \left(\gamma \right)$ the Euler gamma function, $\gamma =\rho +i\eta $, ($\gamma ,\eta \in {\mathbb R}$), and $i=\sqrt{-1} $. 

It has been shown that the PSD and the correlation function are expressed in terms of FSMs as 
\begin{equation} \label{eq16} 
R_{V} \left(\tau \right)=\frac{1}{2\pi i} \int _{\rho -i\infty }^{\rho +i\infty }\nu \left(\gamma \right)\Lambda _{V} \left(-\gamma \right)\left|\tau \right|^{-\gamma } d\gamma   
\end{equation} 
and
\begin{equation} \label{eq17} 
S_{V} \left(\omega \right)=\frac{1}{4\pi i} \int _{\rho -i\infty }^{\rho +i\infty }\Lambda _{V} \left(-\gamma \right)\left|\omega \right|^{\gamma -1} d\gamma   
\end{equation} 

Both integrals are valid for $\rho $ belonging to an interval that depends on the convergence conditions of eq.(\ref{eq14}), which simply becomes the interval $ 0< \rho <1 $ for absolute integrable functions. These equations are a form of the generalized Taylor expansion for functions which are both symmetric and Fourier pairs. Not-symmetric functions are treated in the paper Cottone and Di Paola, 2009a.
Eqs.(\ref{eq16}) and (\ref{eq17}) are integrals along an axis that is parallel to the imaginary axis, called Bromwich's path, and can be simply approximated. Indeed, it suffices to truncate the integration in the interval $\left[-\eta_s ,\eta_s \right]$ and to evaluate the integrand in $2m+1$ points located at steps $\Delta \eta_s =\eta_s/m$. Then, posing $\gamma _{k} =\rho +i\, k\Delta \eta $, with $k=-m,...,m$, the approximated forms of eqs.(\ref{eq16}) and (\ref{eq17}) are
\begin{equation} \label{eq18} 
R_{V} \left(\tau \right)\cong \frac{\Delta \eta }{2\pi } \sum _{k=-m}^{m}\nu \left(\gamma _{k} \right)\Lambda _{V} \left(-\gamma _{k} \right)\left|\tau \right|^{-\gamma _{k} }   
\end{equation} 
and
\begin{equation} \label{eq19} 
S_{V} \left(\omega \right)=\frac{\Delta \eta }{4\pi } \sum _{k=-m}^{m}\Lambda _{V} \left(-\gamma _{k} \right)\left|\omega \right|^{\gamma _{k} -1}   
\end{equation} 
The following simple example clarifies the application of the previous formula.

\subsubsection{Example}
Let us consider a simpler form of eq.(\ref{eq3}) 
\begin{equation} \label{eq20} 
S_{V} \left(\omega \right)=\frac{a}{\left[1+b\, \left|\omega \right|\right]^{5/3} }  
\end{equation} 
with $a>0$ and $b>0$. By using Wolfram's Mathematica software it is easy to calculate FSMs by applying the definition reported in eq.(\ref{eq14}), thus obtaining
\begin{equation} \label{eq21} 
\Lambda _{V} \left(\gamma \right)=\frac{2ab^{-(1+\gamma )} \Gamma \left(2/3-\gamma \right)\Gamma \left(1+\gamma \right)}{\Gamma \left(5/3\right)}  
\end{equation} 
only if $-1<\rho <2/3$, having care that in the whole paper $\rho $ indicates the real part of the complex variable $\gamma $. As in eqs.(\ref{eq16}) and (\ref{eq17}) we use $\Lambda _{V} \left(-\gamma \right)$, the fundamental strip in which such equations are valid is $-2/3<\rho <1$. Inside this interval, we can choose any value of $\rho $ in order to reconstruct the functions $S_{V} \left(\omega \right)$and $R_{V} \left(\tau \right)$, because the integrands are holomorphic. Outside this interval, the residue theorem must be applied. This choice can be profitable and it has been used in Cottone and Di Paola, 2010a to regularize the approximated formula in eqs.(\ref{eq18}) and (\ref{eq19}) in zero. 
\begin{figure}[htbp]
\centering
\begin{minipage}[t]{0.40\linewidth}
\includegraphics[scale =.16]{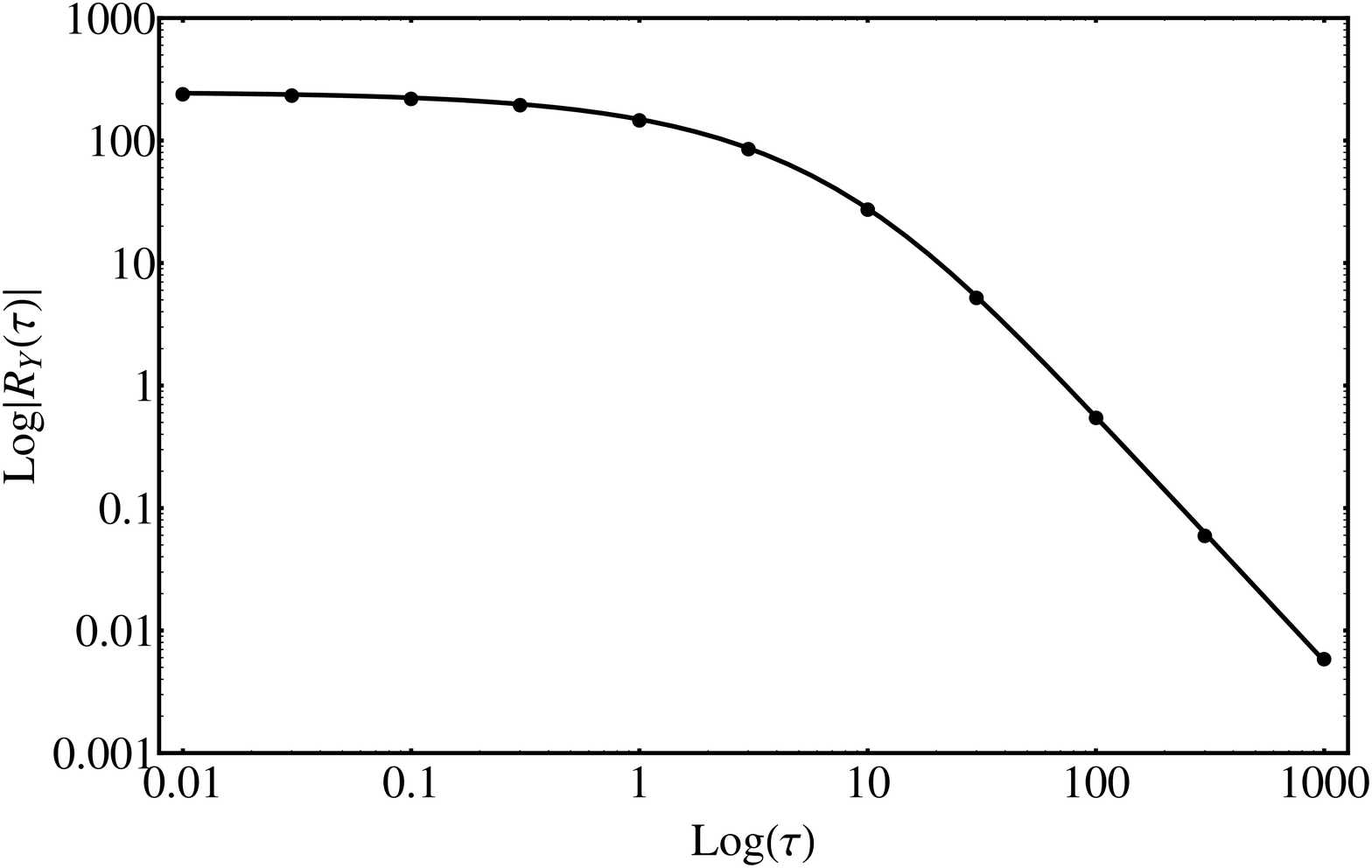}
\end{minipage}\hspace{1.5 cm}
\begin{minipage}[t]{0.45\linewidth}
\includegraphics[scale =.16]{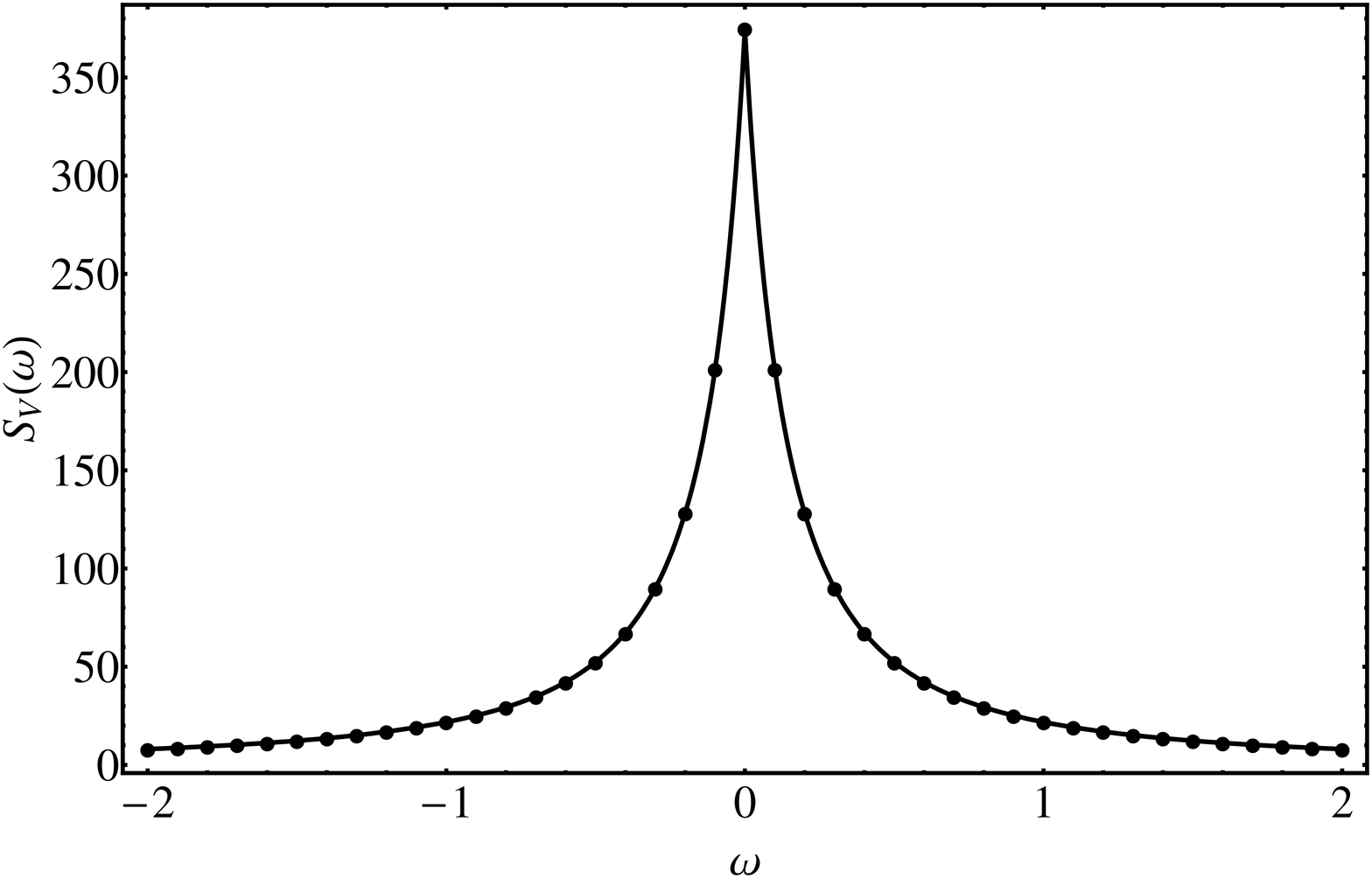}
\end{minipage}
\caption{Comparison between the exact (continuous line) and the approximated (dotted line) correlation function, left panel in log-log diagram, and between the exact (continuous line) and the approximated (dotted line) PSD, right panel}%
\label{Fig1}
\end{figure}

Figures 1 shows the application of eq.(\ref{eq18}) and (\ref{eq19}) with $a=374.8$ and $b=4.51$ which corresponds to having chosen the parameters $z_{0} =0.7$, $u_{*} =2{\rm m/s}$, $\beta =4.96$,$C_{z} =10$, $z=5{\rm m}$ and $L_{V} \left(z\right)=300\left(z/200\right)^{0.67+0.05\ln \left(z_{0} \right)} $in eq.(\ref{eq3}). Sixty FSMs of the type $\Lambda _{V} \left(-\gamma _{k} \right)=1/2+ik\left(0.1\right)$ with $k=-30,...,30$ have been calculated and stored from eq.(\ref{eq21}) and introduced in eqs.(\ref{eq18}) and (\ref{eq19}).
 
This simple example shows that a finite numbed of FSMs can be used to represent the statistics of the Gaussian stationary process $V(t)$. On this solid ground, we can simulate the process $V(t)$ by a fractional linear filter whose coefficients are FSMs.

\section{Simulation of stationary Gaussian stochastic processes by FSMs}
Objective of this section is to represent a Gaussian stationary process $V(t)$ with assigned PSD $S_V(\omega)$ as the output of a fractional differential equation. As shown in Cottone et al. 2010, to this aim, it suffices to consider an ideal linear system ${\cal{L}}(V(t)) = W(t)$ where ${\cal{L}}\left(\cdot \right)$ is a linear operator and $W(t)$ is a Gaussian white noise process with zero mean and correlation function $E[W(t)W(t+\tau )]=q \delta(t)$, and power spectral density $S_{W} =q/(2\pi )$, being $q$  the intensity parameter. From linear system theory, it is known that the output $V(t)$ can be characterized both by the impulse response function $h(t)$ through the Duhamel integral 
\begin{equation} \label{22)} 
V\left(t\right)=\int _{-\infty }^{t} h\left(t-\tau \right)W\left(\tau \right){\rm d}\tau  
\end{equation} 
or by its Fourier transform $H(\omega )$, namely the transfer function, through the input-output relation that reads
\begin{equation} \label{eq23} 
S_{V} \left(\omega \right)=\left|H\left(\omega \right)\right|^{2} S_{W} \left(\omega \right)=\frac{q}{2\pi } \left|H\left(\omega \right)\right|^{2}  
\end{equation} 
Assuming $Arg\left(H\left(\omega \right)\right)=0$, we get a non causal differential equation characterized by the transfer function 
\begin{equation} \label{24)} 
H\left(\omega \right)=|H(\omega )|=\sqrt{\frac{2\pi }{q} S_{V} \left(\omega \right)}  
\end{equation} 

We calculate now the fractional spectral moments of $H(\omega)$ as 
\begin{equation} \label{eq25} 
\Pi _{H} \left(-\gamma \right)\mathop{=}\limits^{def} \int _{-\infty }^{\infty } \left|\omega \right|^{-\gamma } H\left(\omega \right){\rm d}\omega ,{\kern 1pt} {\kern 1pt} {\kern 1pt} {\kern 1pt} {\kern 1pt} {\kern 1pt} {\kern 1pt} {\kern 1pt} {\kern 1pt} {\kern 1pt} {\rm R}e\gamma >0 
\end{equation} 
that are called \textit{H-Fractional Spectral Moments} (H-FSMs). As before, it can be shown that such a function can be used to represent both the impulse response in the time domain as 
\begin{equation} \label{26)} 
h\left(t\right)=\frac{1}{\left(2\pi \right)^{2} i} \int _{\rho -i\infty }^{\rho +i\infty } \nu \left(\gamma \right)\Pi _{H} \left(-\gamma \right)t^{-\gamma } {\rm d}\gamma ,{\kern 1pt} {\kern 1pt} {\kern 1pt} {\kern 1pt} {\kern 1pt} {\kern 1pt} {\kern 1pt} {\kern 1pt} {\kern 1pt} {\kern 1pt} {\kern 1pt} {\kern 1pt} {\kern 1pt} t>0
\end{equation} 
and the transfer function in the form 
\begin{equation} \label{27)} 
H\left(\omega \right)=\frac{1}{4\pi i} \int _{\rho -i\infty }^{\rho +i\infty } \Pi _{H} \left(-\gamma \right)\left|\omega \right|^{\gamma -1} {\rm d}\gamma 
\end{equation} 

Following the same notation of the previous section, approximated form of the last integrals are
\begin{equation} \label{28)} 
h\left(t\right)\cong \frac{\Delta \eta }{\left(2\pi \right)^{2} } \sum _{k=-m}^{m} \nu \left(\gamma _{k} \right)\Pi _{H} \left(-\gamma _{k} \right)t^{-\gamma _{k} }  
\end{equation} 

\begin{equation} \label{29)} 
H\left(\omega \right)\cong \frac{\Delta \eta }{4\pi } \sum _{k=-m}^{m} \Pi _{H} \left(-\gamma _{k} \right)\left|\omega \right|^{\gamma _{k} -1}  
\end{equation} 
As shown in Cottone et al. 2010, the input-output relation for linear system and Fourier transform properties of Riesz fractional operators lead to the relevant representation of the Gaussian stationary process
\begin{equation} \label{eq30} 
V\left(t\right)=\frac{1}{4\pi i} \int _{\rho -i\infty }^{\rho +i\infty } \Pi _{H} \left(-\gamma \right)\left(I^{1-\gamma } W\right)\left(t\right){\rm d}\gamma  
\end{equation} 
whose PSD is $S_{V} (\omega )$. The approximated form reads
\begin{equation} \label{eq31} 
V\left(t\right)=\frac{\Delta \eta }{4\pi } \sum _{k=-m}^{m} \Pi _{H} \left(-\gamma _{k} \right)\left(I^{1-\gamma _{k} } W\right)\left(t\right) 
\end{equation} 
and readers must bear in mind that this approximation carries out a truncation and discretization error, that can be made arbitrarily small. Eq.(\ref{eq30}), or its discretized counterpart given in eq.(\ref{eq31}), show that the process $V(t)$ may be obtained as the superposition of the fractional integrals of a Gaussian white noise process. 

\subsection{Implementation}
We continue the previous example by simulating a digital temporal signal by means of eq.(\ref{eq31}). FSMs have been already calculated and, for the functional form in eq.(\ref{eq20}) they are given by eq.(\ref{eq21}). It is reasonable to keep the parameters $\rho =1/2$, ${\rm \Delta }\eta =0.1$ and $m=30$ by which, in the previous example, we achieved to represent both the PSD and the correlation function, to a good accuracy. Firstly, we must calculate the H-FSMs, and this task can be strongly simplified by using of the Wolfram's Mathematica, which returns
\begin{equation} \label{32)} 
\Pi _{H} \left(\gamma \right)=\frac{2\sqrt{2\pi a} b^{-(1+\gamma )} \Gamma \left(-1/6-\gamma \right)\Gamma \left(1+\gamma \right)}{\Gamma \left(5/6\right)}  
\end{equation} 
Then, the practical implementation of eq.(\ref{eq31}) requires a further step to calculate the Riesz fractional integral of the Gaussian white noise $W\left(t\right)$. To this aim, we firstly exploit the equivalence between the Riesz's fractional integral and the Grünwald -- Letnikov's discrete fractional operator (see Samko et al. 1993) that in approximated form is expressed as
\begin{equation} \label{eq33} 
\left(I^{\gamma } W\right)\left(t\right)\cong \frac{\tau^{\gamma } }{2{\rm cos}\left(\gamma \pi /2\right)} \left[\left(\Delta _{+}^{-\gamma } W\right)\left(t\right)+\left(\Delta _{-}^{-\gamma } W\right)\left(t\right)\right] 
\end{equation} 
where
\begin{equation} \label{34)} 
\left(\Delta _{\pm }^{\gamma } W\right)\left(t\right)=\sum _{k=0}^{\infty }\left(-1\right)^{k} \left(\begin{array}{l} {\gamma } \\ {k} \end{array}\right)W\left(t\mp k \tau\right)  
\end{equation} 
being $\tau$ a discrete time step. To the limit $\tau\to 0$, eq.(\ref{eq33}) becomes an identity. These relations are true on the whole real support and they can be also rewritten in the case of a finite interval. Let us consider a time window $\left[0,T\right]$ partitioned into $n$ time steps of amplitude $\tau$. At each node $0,1,...,n$, the noise $W\left(0\right),W\left(h\right),...,W\left(j \tau\right),...,W\left(n \tau\right)$ is evaluated as a realization of a Gaussian random variable with zero mean and variance $\sqrt{\tau} $, having chosen $q=1$. The Riesz's fractional integral of such a noise is calculated by the Grünwald - Letnikov's approach as
\begin{equation} \label{eq35} 
\left(I^{\gamma } W\right)\left(j \tau\right)\cong \frac{\tau^{-\gamma } }{2{\rm cos}\left(\gamma \pi /2\right)} \left[\sum _{k=0}^{j}\left(-1\right)^{k} \left(\begin{array}{l} {-\gamma } \\ {k} \end{array}\right)W\left(j \tau-k \tau\right) +\sum _{k=0}^{n-j}\left(-1\right)^{k} \left(\begin{array}{l} {-\gamma } \\ {k} \end{array}\right)W\left(j \tau+k \tau\right) \right] 
\end{equation} 
The coefficients in the latter can be efficiently calculated both iteratively and by using the Fast Fourier transform as reported in Podlubny, p.209. Such coefficients, calculated for ${\rm Re}\gamma >-1$, decrease with inverse power-law behavior and, for many functions, they can be neglected after a finite number of terms, say $p$. In the case here considered, with $\rho =1/2$, $p=400$ terms suffice to achieve the searched accuracy. In other words, the fractional integral of the Gaussian white noise at the time step  $j \tau$ is calculated by eq.(\ref{eq35}) considering the influence of $p$ noise realizations which precede and follow the instant $j$.  

\begin{figure}[h]
\centering
\includegraphics[scale =.3]{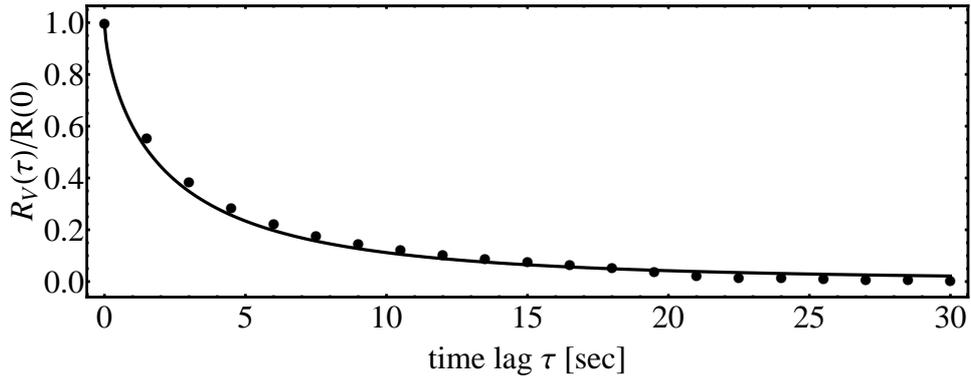}
\caption{Comparison between the exact (continuous line) and the simulated (dotted line) correlation function.}
\label{fig2}
\end{figure}

Figure 2 shows the comparison between the target correlation function and the correlation function of the time series $V\left(j \tau\right)$ with $j=0,...,n$. The simulation has been implemented in Octave, see Eaton, 2002, with the parameters: $3\cdot 10^{6} $ time steps and $\tau=5\cdot 10^{-2}$.

\section{Extension to multivariate stochastic processes}
Let us consider a vector of Gaussian white noise processes, ${\bf{W}}(t)$, with zero mean and diagonal power spectral density matrix $\bf{S_{W}} (\omega)$, whose elements along the diagonal are $S_{W_{j} W_{j} } \left(\omega \right)=q_{j} /\left(2\pi \right)$, where $q_{j} $ is the strength of the jth noise component. We can model ${\bf{V}}\left(t\right)$ as the output of a multidimensional linear system characterized by the impulse response matrix ${\bf{H}}(t)$ that can be expressed by Duhamel's integral 
\begin{equation} \label{37)} 
{\bf{V}}\left(t\right)=\int _{-\infty }^{t} {\bf{H}}\left(t-s \right){\bf{W}}\left(s \right){\rm d}s  
\end{equation} 
Indicating by ${\bf{S}}_{\bf{V}} \left(\omega \right)$ the power spectral density matrix of the output, from the linearity of the system it follows that 
\begin{equation} \label{eq38} 
{\bf{S}}_{\bf{V}} \left(\omega \right)={\bf{H}}\left(\omega \right)\,{\bf{S}}_{\bf{W}}^{1/2} \left(\omega \right)\, {\bf{S}}_{\bf{W}}^{1/2} \left(\omega \right)\bar{{\bf{H}}}^{T} \left(\omega \right) 
\end{equation} 
where ${\bf{H}}\left(\omega \right)$ is the transform matrix and the bar means conjugation. Rewriting eq.(\ref{eq6}) as 
\begin{equation} \label{39)} 
{\bf{S}}_{{\bf{V}}} \left(\omega \right)={\bf{\Psi}} \left(\omega \right){\bf{L}} ^{1/2} \left(\omega \right)\, {\bf{L}} ^{1/2} \left(\omega \right)\bar{{\bf{\Psi}} }^{T} \left(\omega \right) 
\end{equation} 
and comparing the latter with eq.(\ref{eq38}) the characterization of the linear system can be obtained by selecting the transfer function as
\begin{equation} \label{40)} 
{\bf{H}}\left(\omega \right)={\bf{\Psi}}\left(\omega \right)\, {\bf{L}} ^{1/2} \left(\omega \right) 
\end{equation} 
with the choice of $q_{j} =2\pi $ for every noise component $W_{j} \left(t\right)$.The output of course will be Gaussian due to the linearity of the system. 

The next steps we want to take aims to represent ${\bf{V}}\left(t\right)$ in terms of fractional spectral moments of the transfer matrix. This will turn out to have relevant computational advantages. Firstly we will introduce the fractional moments of the transfer matrix ${\bf{H}}(\omega )$; then we will represent ${\bf{H}}(\omega )$ as a sum of fractional spectral moments and finally we will give the expression of the process ${\bf{V}}\left(t\right)$ with assigned ${\bf{S}}_{\bf{V}}\left(\omega \right)$ in terms of the fractional spectral moments of ${\bf{H}}(\omega )$. 

Let us define the H-FSMs of ${\bf{H}}(\omega )$ of complex order $\gamma =\rho +i\eta $, labeled in the following by ${\bf{\Pi}}\left(\gamma \right)$, as 
\begin{equation} \label{eq41} 
{\bf{\Pi}} \left(\gamma \right)\mathop{=}\limits^{def} \int _{-\infty }^{\infty } \left|\omega \right|^{\gamma } {\bf{H}}\left(\omega \right){\rm d}\omega  
\end{equation} 
with $\rho $ chosen such that the integral converges. 

In perfect analogy to the eqs.(\ref{28)}) and (\ref{29)}) we can obtain the representation of the impulse response matrix ${\bf{H}}(t)$ in the time domain as 
\begin{equation} \label{eq43} 
{\bf{H}}\left(t\right)=\frac{1}{\left(2\pi \right)^{2} i} \int _{\rho -i\infty }^{\rho +i\infty } \nu \left(\gamma \right){\bf{\Pi}} \left(-\gamma \right)t^{-\gamma } {\rm d}\gamma \end{equation} 
and the transfer function ${\bf{H}}(\omega)$ in the form 
\begin{equation} \label{eq44} 
{\bf{H}}\left(\omega \right)=\frac{1}{4\pi i} \int _{\rho -i\infty }^{\rho +i\infty } {\bf{\Pi}} \left(-\gamma \right)\left|\omega \right|^{\gamma -1} {\rm d}\gamma 
\end{equation} 

Assuming that all the components of the transfer matrix ${\bf{H}}\left(\omega \right)$ are absolute integrable, in the following we will consider the fundamental strip of eqs.(\ref{eq43}) and (\ref{eq44}) such as  $0<\rho <1$. 

A simple approximation of the latter relations can be given by truncating the integral along the imaginary axis $\eta $. Indicating as $\eta_s$ the truncation limit for the integrals, let us divide the interval $\left[-\eta_s, \eta_s\right]$ in $2m$ intervals of amplitude $\Delta \eta =\eta_s /m$, with $m\in {\rm {\mathbb N}}$. Then, the integrals in eqs.(\ref{eq43}) and (\ref{eq44}) can be approximated by the values at the nodes $\gamma _{k} =\rho +{\rm i}k\Delta \eta $ in the form 
\begin{equation} \label{eq45} 
H_{rs} \left(t\right)\cong \frac{\Delta \eta }{\left(2\pi \right)^{2} } \sum _{k=-m}^{m} \nu \left(\gamma _{k} \right)\Pi _{rs} \left(-\gamma _{k} \right)t^{-\gamma _{k} }  
\end{equation} 
and 
\begin{equation} \label{eq46} 
H_{rs} \left(\omega \right)\cong \frac{\Delta \eta }{4\pi } \sum _{k=-m}^{m} \Pi _{rs} \left(-\gamma _{k} \right)\left|\omega \right|^{\gamma _{k} -1}  
\end{equation} 

The latter equations hold true for each component $r,s = 1, ..., N$ of the matrices ${\bf{H}}(t)$ and ${\bf{H}}(\omega)$.

Now, having represented the transfer function both in exact and approximated form in terms of H-FSM we are ready to infer an analytic expression for the representation of the stationary Gaussian vector process ${\bf{V}}(t)$.

First, the input-output relation of eq.(\ref{32)}) is written in Fourier domain as ${\bf{V}}\left(\omega ,T\right)={\bf{H}}\left(\omega \right){\bf{W}}\left(\omega ,T\right)$, where $T>0$ is a truncation bound. Then, by introducing the representation of the transfer matrix in terms of H-FSMs, one obtains 
\begin{equation} \label{47)} 
{\bf{V}}\left(\omega ,T\right)=\frac{1}{4\pi i} \int _{\rho -i\infty }^{\rho +i\infty } {\bf{\Pi}} \left(-\gamma \right)\left|\omega \right|^{\gamma -1} {\bf{W}}\left(\omega ,T\right){\rm d}\gamma  
\end{equation} 
and by inverse Fourier transform, we finelly obtain the relation searched that reads 
\begin{equation} \label{eq48} 
{\bf{V}}\left(t\right)=\frac{1}{4\pi i} \int _{\rho -i\infty }^{\rho +i\infty } {\bf{\Pi}} \left(-\gamma \right)\left(I^{1-\gamma } {\bf{W}}\right)\left(t\right)d\gamma  
\end{equation} 

In the previous derivation we have used the property of the inverse Fourier transform of the Riesz's fractional integrals 
\[\mathop{\lim }\limits_{T\to \infty } F^{-1} \left\{\left|\omega \right|^{\gamma -1} W_{j} \left(\omega ,T\right);t\right\}=\left(I^{1-\gamma } W_{j} \right)\left(t\right)\] 
that is true for each component of ${\bf{W}}\left(t\right)$.

Eq.(\ref{eq48}) is a novel exact representation of the multivariate stationary Gaussian process with assigned PSD matrix, in which the term 
\begin{equation} \label{49)} 
\left(I^{1-\gamma } {\bf{W}}\right)\left(t\right)=\left[\begin{array}{cccc} {I^{1-\gamma } W_{1} } & {I^{1-\gamma } W_{2} } & {...} & {I^{1-\gamma } W_{N} } \end{array}\right]^{T}  
\end{equation} 
is a vector whose elements are the Riesz fractional integrals of complex order $1-\gamma $ of the independent white noise processes $W_{j} $, $j=1,2,...,N$. The fundamental strip in which Eq.(\ref{eq48}) converges, depends on the integrability of the single components of Eq.(\ref{eq41}). Following our assumption that the components of the matrix ${\bf{H}}\left(\omega \right)$ are absolute integrable, the integral in Eq.(\ref{eq48}) can be performed inside the strip $0<\rho <1$. As already pointed out, any values of $\rho $ inside this fundamental strip can be chosen.

\subsection{ Approximated decomposition by H-FSM}
Eq.(\ref{eq48}) is approximated in a quite simple but effective way, following the same trace of the approximations in eqs.(\ref{eq45}) and (\ref{eq46}). To this aim, a value of $\rho $ is chosen inside the interval $0<\rho <1$. Then, the integrator is rewritten as $d\gamma =id\eta $ and the bounds of integration are truncated up to the values $\left[-\eta_s ,\eta_s \right]$. In this way we are introducing a truncation error in the evaluation of ${\bf{V}}(t)$, and the integral becomes
\begin{equation} \label{eq50} 
{\bf{V}}\left(t\right)\cong \frac{1}{4\pi } \int _{\eta_s }^{\eta_s} {\bf{\Pi}}\left(-\gamma \right)\left(I^{1-\gamma } {\bf{W}}\right)\left(t\right)d\eta  
\end{equation} 

The interval $\left[-\eta_s, \eta_s \right]$ is partitioned in $2m$ intervals of amplitude $\Delta \eta =\eta_s/m$, with $m\in {\rm {\mathbb N}}$ and the integral in eq.(\ref{eq50}) is evaluated at the nodes $\gamma _{k} =\rho +{\rm i}k\Delta \eta $. This correspond to apply a rectangular numerical scheme to the evaluation of eq.(\ref{eq50}), that reads 

\begin{equation} \label{eq51} 
{\bf{V}}\left(t\right)\cong \frac{\Delta \eta }{4\pi } \sum _{k=-m}^{m}{\bf{\Pi}} \left(-\gamma _{k} \right)\left(I^{1-\gamma _{k} }{\bf{W}}\right)\left(t\right)  
\end{equation} 
The accuracy of the latter formula can be increased by using more accurate numerical schemes to approximate the exact decomposition in eq.(\ref{eq48}). This kind of problem is known in the field of the numerical treatment of the Laplace transform inversion, that has the same mathematical structure of the problem here treated. Talbot's contours, conformal mapping, higher numerical integration schemes are of course applicable to our problem, but the simplicity and accuracy of eq.(\ref{eq51}) are satisfactory and no further numerical effort is needed to our scopes. It is important to emphasize that to practically use eq.(\ref{eq51}), $(2m+1)\times N^{2} $numbers, i.e. the H-FSM at different values of $k=-m,...,m$, must be calculated. This number can be strongly reduced by the spectral decomposition of the PSD matrix recalled in section 2.

\subsection{H-FSM in reduced space} 
The computational effort for the calculation of the H-FSM decomposition in eq.(\ref{eq48}) is mainly influenced by the calculation of the fractional spectral moments. A considerable reduction of computational effort can be achieved by considering the eigen-properties of the PSD matrix. Indeed, following the paper of Di Paola and Gullo, 2001, in many applications, like in wind engineering, only a reduced number of terms $M\ll N$ is relevant in simulating the N-variate process ${\bf{V}}(t)$, in the form of eq.(\ref{eq12}). 

Applying this reasoning to the H-FSM decomposition, a reduced form of the integral representation by H-FSMs can be given in the form
\begin{equation} \label{52)} 
{\bf{V}}\left(t\right)\cong \frac{1}{4\pi i} \int _{\rho -i\infty }^{\rho +i\infty } \tilde{{\bf{\Pi}} }\left(-\gamma \right)\left(I^{1-\gamma } \tilde{{\bf{W}}}\right)\left(t\right)d\gamma  
\end{equation} 
where $\tilde{{\bf{\Pi}}}\left(-\gamma \right)$ is a $\left(N\times M\right)$ reduced matrix and $\left(I^{1-\gamma } \tilde{{\bf{W}}}\right)\left(t\right)$ is a $\left(M\times 1\right)$ vector and $\tilde{{\bf{\Pi}} }\left(-\gamma \right)$ is calculated on the reduced matrix ${\bf \tilde H}\left( \omega  \right)$ as
\[
{\bf \tilde H}\left( \omega  \right)_{N \times M}  = {\bf \tilde \Psi }\left( \omega  \right)_{N \times M} {\bf \tilde L}_{M \times M} 
\]

By such a reduction, the discrete approximated form of the latter reads
\begin{equation} \label{53)} 
{\bf{V}}\left(t\right)\cong \frac{\Delta \eta }{4\pi } \sum _{k=-m}^{m}\tilde{{\bf{\Pi}} }\left(-\gamma _{k} \right)\left(I^{1-\gamma _{k} } \tilde{{\bf{W}}}\right)\left(t\right)  
\end{equation} 
and the computation is performed storing $(2m+1)\times N\times M$ numbers, with $M\ll N$.

\section{Conclusions}

We have introduced a novel method for the representation and the consequent digital simulation of stationary Gaussian processes and multivariate fields. The methods is developed introducing the Fractional Spectral Moments. If these features are calculated from the power spectral density function, then they can be used to represent both in exact and in approximate form the correlation and the spectral density, as shown in section 3.
If the Fractional Spectral Moments are calculated by integrating the transfer matrix of a  linear system excited by a Gaussian white noise, then they are the coefficients of the time series given in eq.(\ref{eq31}) which restore the target process. This method can be seen as a valid alternative to classical ARMA models, especially in case of PSD function with pathological behavior, see Cottone and Di Paola, 2010a. 

Moreover the paper extends the use of the fractional spectral moments to represent N-variate Gaussian processes. Two steps are required to this aim. The first step consists in performing a spectral decomposition of the PSD matrix of the N-variate process, retaining only the most relevant $M\ll N$ eigenvectors. Then, such quantities are used to calibrate the transfer matrix of a reduced multidimensional linear system whose output is the searched process and the input are M uncorrelated Gaussian white noise processes with unitary power spectral densities.

\section{References}

Borgman L.E. (1969). Ocean wave simulation for engineering design. \textit{. J. Waterways and Harbour Div. ASCE}, pp.557-583.

Cottone G., Di Paola M., Pirrotta A. (2008). Path integral solution by fractional calculus. Journal of Physics: Conference Series, Vol.96, pp.1-11.

Cottone G., Di Paola M. (2009a). On the use of fractional calculus for the probabilistic characterization of random variables. \textit{Probabilistic Engineering Mechanics}, Vol. 24, pp. 321-330.

Cottone G., Di Paola M., Marino F. (2009b). On the Derivation of the Fokker-Plank Equations by using of Fractional Calculus. \textit{The 10th International Conference on Structural Safety and Reliability ICOSSAR 2009}. Osaka (Japan).

Cottone G., Di Paola M. (2010a).  A New Representation of Power Spectral Density and Correlation Function by Means of Fractional Spectral Moments. \textit{Probabilistic Engineering Mechanics}, Vol. 25, pp. 348-353.

Cottone G., Di Paola M., Santoro, R. (2010b) A novel exact representation of stationary colored Gaussian processes (fractional differential approach).\textit{Journal of Physics A: Mathematical and Theoretical}. Vol.43(8), pp.085002.

Deodatis G. (1995). Simulation of multivariate stochastic processes. in: Spanos (Ed.), \textit{Comp. Stochastic Mechanics}, Balkema, Rotterdam, pp.297-305.

Deodatis G., Shinozuka M. (1988). Autoregressive model for non-stationary stochastic processes. \textit{J. Eng. Mech. ASCE}  Vol.114 (11), pp.1995-2012.

Di Paola M. (1998). Digital simulation of wind field velocity. \textit{Journal of Wind Engineering and Industrial Aerodynamics}. Vol.74-76, pp. 91- 109.

Di Paola M., Gullo I. (2001). Digital generation of multivariate wind field processes, \textit{Probabilistic Engineering Mechanics}. Vol.16, pp. 1-10.

Eaton, J.W. (2002). \textit{GNU Octave Manual.} Network Theory Limited.

Kareem A. (2006). Numerical simulation of wind effects: a probabilistic perspective. \textit{The Fourth International Symposium on Computational Wind Engineering} (CWE2006), Yokohama.

Kozin F. (1988). Autoregressive moving average models of earthquake records. \textit{ Probabilistic Engineering Mechanics} Vol.3(2), pp.58-63.

Naganuma T., Deodatis G., Shinozuka M. (1987). ARMA model for two dimensional processes. \textit{J. Eng. Mech. Div. ASCE} Vol. 113 (2), pp.234-251.

Podlubny, I. (1998). \textit{Fractional Differential Equations: An Introduction to Fractional Derivatives, Fractional Differential Equations, to Methods of Their Solution and Some of Their Applications}. Mathematics in Science and Engineering, Elsevier.

Samko S., Kilbas A.A. and Marichev O.I. (1993) \textit{Fractional Integrals and Derivatives. Theory and Applications}. Gordon and Breach Science Publishers, Switzerland.

Saramas E., Shinozuka M., Tsurui (1985). ARMA representation of random processes, \textit{J. Eng. Mech. ASCE} Vol. III(3), pp.449-461.

Shinozuka M. (1971).  Simulation of multivariate and multidimensional random processes. \textit{ J. of Acoustical Society of America}, Vol. 49, pp.357-367.

Simiu E., Scanlan R. (1996). \textit{Wind effects on Structures}. John Wiley \& Sons, New York

Solari G., Piccardo G. (2001). Probabilistic 3-D turbulence modelling for gust buffeting of structures. \textit{Probababilistic Engineering Mechanics} Vol.16, pp.73-86.

Spanos P.D., Mignolet M.P. (1986). Z-transform modeling of P-M wave spectrum. \textit{ J. Eng. Mech. Div. ASCE} Vol.192(8), pp.745-759.

Vanmarcke E. (1972). Properties of spectral moments with applications to random vibrations. \textit{Journal of Engineering Mechanical Division, ASME}, Vol. 42, pp.215-20.

\end{document}